\documentclass[aps,pre,preprint,superscriptaddress]{revtex4-1}
\usepackage{latexsym,amssymb,amsmath,amsfonts,amsthm}

\usepackage{color}

\def\beq{ \begin{equation} }
\def\eeq{ \end{equation} }
\def\mn{\medskip\noindent}
\def\ms{\medskip}

\def\square{\vcenter{\vbox{\hrule height .4pt
  \hbox{\vrule width .4pt height 5pt \kern 5pt
        \vrule width .4pt} \hrule height .4pt}}}

\usepackage{graphicx}

\begin{document}

\title{A multi-opinion evolving voter model with infinitely many phase transitions}
\author{Feng Shi}
\author{Peter J. Mucha}
\affiliation{Dept.~of Mathematics, CB\#3250, U. of North Carolina, Chapel Hill, NC 27599-3250}
\author{Rick Durrett}
\affiliation{Dept.~of Mathematics, Box 90320, Duke University, Durham, NC 27708-0320}

\begin{abstract}
  We consider an idealized model in which individuals' changing opinions and their social network coevolve, with disagreements between neighbors in the network resolved either through one imitating the opinion of the other or by reassignment of the discordant edge. Specifically, an interaction between $x$ and one of its neighbors $y$ leads to $x$ imitating $y$ with probability $(1-\alpha)$ and otherwise (i.e., with probability $\alpha$) $x$ cutting its tie to $y$ in order to instead connect to a randomly chosen individual. Building on previous work about the two-opinion case, we study the multiple-opinion situation, finding that the model has infinitely many phase transitions. Moreover, the formulas describing the end states of these processes are remarkably simple when expressed as a function of $\beta = \alpha/(1-\alpha)$.
\end{abstract}

\maketitle

\section{Introduction}

In the last decade, there have been a number of studies of systems in which the states of individuals and the connections between them coevolve, see \cite{GB,GS}. The systems considered include evolutionary games \cite{EB}--\cite{PTM} and epidemics \cite{GDB}--\cite{SSP}, but here we will concentrate on the spread of opinions \cite{GZ}--\cite{IKKB}. Different from the models of cascades \cite{G}--\cite{GC} which are also widely used in the study of opinion spread, the evolving voter model we study here allows an agent to switch between different opinions and the network topology to change accordingly, yet we assume that agents impose equal influence over each other (cf., multi-state complex contagions \cite{BL}--\cite{MWGP}). This model provides building blocks to quantitatively study collective behaviors in various social systems, e.g., segregation of a population into two or more communities with different political opinions, religious beliefs, cultural traits, etc.

We are particularly interested here in systems that generalize the model proposed by Holme and Newman \cite{HN}. In their model there is a network of $N$ vertices and $M$ edges. The individual at vertex $v$ has an opinion $\xi(v)$ from a set of $G$ possible opinions and the number of people per opinion $\gamma_N = N/G$ stays bounded as $N$ gets large. On each step of the process, a vertex $x$ is picked at random. If its degree $d(x)$ equals $0$, nothing happens. If $d(x)>0$, (i) then with probability $1-\alpha$ a random neighbor $y$ of $x$ is selected and we set $\xi(x)=\xi(y)$; (ii) otherwise (i.e., with probability $\alpha$) an edge attached to vertex $x$ is selected and the other end of that edge is moved to a vertex chosen at random from those with opinion $\xi(x)$.  This process continues until the `consensus time' $\tau$, at which there are no longer any discordant edges---that is, there are no edges connecting individuals with different opinions.

For $\alpha=1$, only rewiring steps occur, so once all of the $M$ edges have been touched, the graph has been disconnected into $G$ components, each consisting of individuals who share the same opinion. Since none of the opinions have changed, the components of the final graph are all small (i.e., their sizes are Poisson with mean $\gamma_N$). By classical results for the coupon collector's problem, this requires $\sim M \log M$ updates, see e.g., page 57 in \cite{D10}. In the case of sparse graphs we consider here $M\sim cN$ (i.e., $M/N \to c$) so the number of steps is $O(N \log N)$, i.e., when $N$ is large it will be $\approx C N \log N$.

In contrast, for $\alpha=0$ this system reduces to the voter model on a static graph. If we suppose that the initial graph is an Erd\H{o}s-R\'enyi random graph in which each vertex has average degree $\lambda>1$, then (see e.g., Chapter 2 of \cite{D08}) there is a ``giant component'' that contains a positive fraction, $\mu$, of the vertices and the second largest component is small having only $O(\log N)$ vertices. The voter model on the giant component will reach consensus in $O(N^2)$ steps (see, e.g., Section 6.9 of \cite{D08}), so the end result is that one opinion has $\mu N$ followers while all of the other groups are small.

Using simulation and finite size scaling, Holme and Newman showed that there is a critical value $\alpha_c$ so that for $\alpha > \alpha_c$ all of the opinions have a small number of followers at the end of the process, while for $\alpha<\alpha_c$ ``a giant community of like-minded individuals forms.'' When the average degree $\lambda=2M/N=4$ and the number of individuals per opinion $\gamma_N\to10$, this transition occurs at $\alpha_c \approx 0.46$. See \cite{KH}--\cite{BG2} for recent work on this model.

In \cite{evo8}, we studied a two-opinion version of this model in which on each step an edge is chosen at random and is given a random orientation, $(x,y)$. If the individuals at the two ends have the same opinion nothing happens. If they differ, then (i) with probability $1-\alpha$ we set $\xi(x)=\xi(y)$; (ii) otherwise (i.e., with probability $\alpha$) $x$ breaks its edge to $y$ and reconnects to (a) a vertex chosen at random from those with opinion $\xi(x)$, a process we label `rewire-to-same', or (b) at random from the graph, a process we label `rewire-to-random'. Here, we will concentrate on the second rewiring option, rewire-to-random. While this process may be less intuitive than the rewire-to-same version, it has a more interesting phrase-transition, as documented in \cite{evo8}.

The remainder of this paper is organized as follows. In Section \ref{sec:2}, we recall the main results from \cite{evo8} that provide essential context for our observations of the multiple-opinion case, which we begin to explore in Section \ref{sec:multi}. We then continue in Section \ref{sec:quant} with further quantitative details about the phase transitions and their underlying quasi-stationary distributions, before concluding comments in Section \ref{sec:conclusion}.

\section{Two-opinion model}
\label{sec:2}

Suppose, for concreteness, that the initial social network is an Erd\H{o}s-R\'enyi random graph in which each individual has average degree $\lambda >1$, and that vertices are assigned opinions 1 and 0 independently with probabilities $u$ and $1-u$. Simulations suggest that the system has the following

\mn
{\bf Phase transition.} {\it For each initial density $u \le 1/2$ there is a critical value $\alpha_c(u)$ so that for $\alpha > \alpha_c(u)$, consensus occurs after $O(N \log N)$ updates and the fraction of voters with the minority opinion at the end is $\rho(\alpha,u) \approx u$. For $\alpha < \alpha_c(u)$ consensus is slow, requiring $O(N^2)$ updates, and $\rho(\alpha,u) \approx \rho(\alpha,0.5)$.}

\begin{figure} 
\centering
  \includegraphics[width=0.45\textwidth]{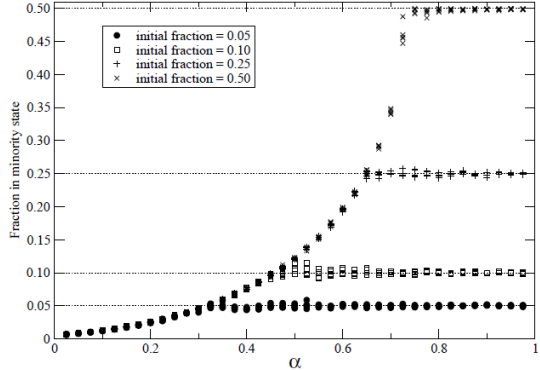}
  \caption{Simulation results for the rewire-to-random model, starting from Erd\H{o}s-R\'{e}nyi graphs with N=100,000 nodes and average degree $\lambda=4$ \cite{evo8}. The final fractions of the minority below phase transitions follow an universal curve independent of the initial fractions.}
  \label{fig:rewire_to_random_P1}
\end{figure}

\ms
To help understand the last statement, the reader should consult the picture in Figure \ref{fig:rewire_to_random_P1}. If the initial fraction of 1's $u=1/2$ then as $\alpha$ decreases from 1, the ending density $\rho(\alpha,1/2)$ stays constant at 1/2 until $\alpha=\alpha_c(1/2)$ and then decreases to a value close to 0 at $\alpha=0$. For convenience, we call the graph of $\rho(\alpha,1/2)$ for $\alpha<\alpha_c\doteq 0.74$, the {\it universal curve}. If the initial density is $u<1/2$, then the ending density $\rho(\alpha,u)$ stays constant at $u$ until the flat line $(\alpha,u)$ hits the universal curve and then $\rho(\alpha,u) \approx \rho(\alpha,0.5)$ for $\alpha<\alpha_c(u)$. The main aim of \cite{evo8} was to use simulations, heuristic arguments, and approximate models to explain the presence and properties of this universal curve describing the consensus states that result from the slow-consensus process. To make it easier to compare the results here with the previous paper, we rescale time so that times between updating steps are exponential with rate $M$, where $M$ is the total number of edges.

\mn
{\bf Quasi-stationary distributions.} {\it
Let $v(\alpha)=\rho(\alpha,0.5)$. If $\alpha < \alpha_c(1/2)$ and $v(\alpha) < u \le 1/2$ then starting from product measure with density $u$ of 1's, the evolving voter model converges rapidly to a quasi-stationary distribution $\nu_{\alpha,u}$. At time $tM$ the evolving voter model looks locally like $\nu_{\alpha,\theta(t)}$ where the density changes according to a generalized Wright-Fisher diffusion process
\beq
d\theta_t = \sqrt{(1-\alpha) [c_\alpha\theta_t (1-\theta_t) - b_\alpha]} dB_t
\label{WF}
\eeq
until $\theta_t$ reaches $v(\alpha)$ or $1-v(\alpha)$, the two solutions of $c_\alpha x(1-x) = b_\alpha$.}

\begin{figure} 
  \centering
  \includegraphics[width=0.5\textwidth]{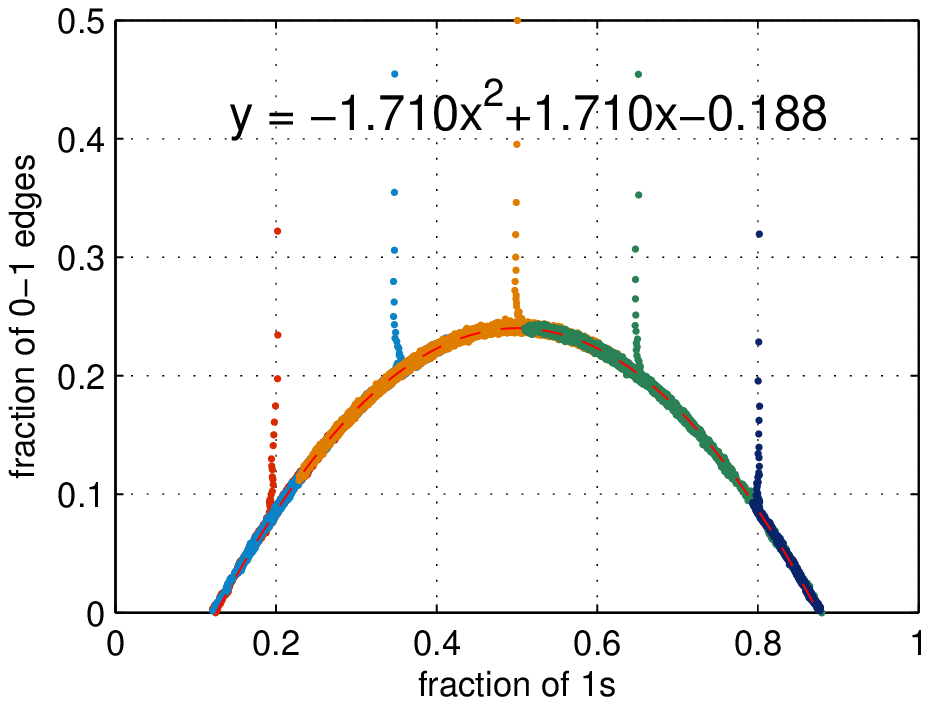}
  \includegraphics[width=0.5\textwidth]{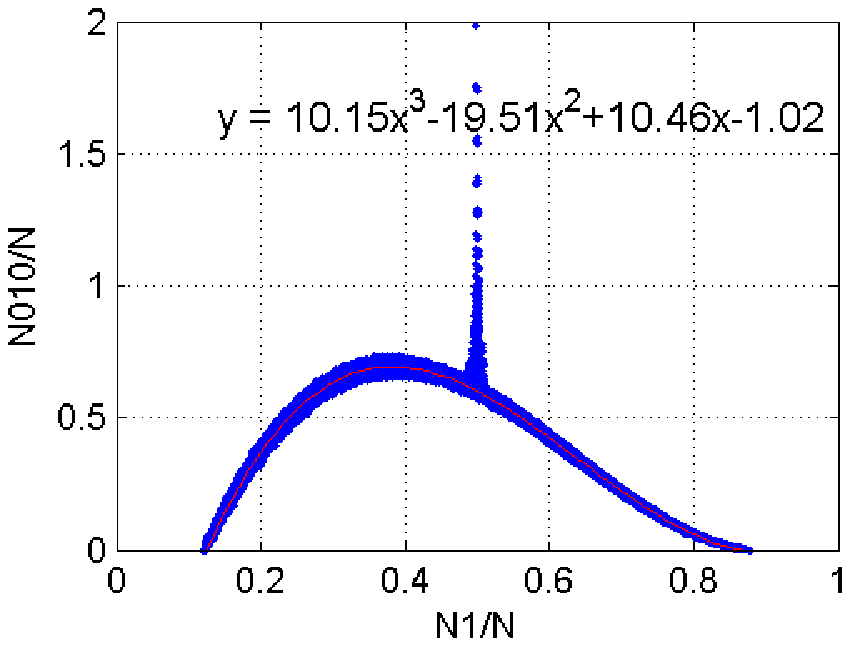}
  \caption{(Top) Evolution of the fraction of edges that are discordant $0$-$1$ edges, $N_{01}(t)/M$, versus the population of opinions $N_1(t)/N$ when $\alpha=0.5$ for the rewire-to-random dynamic. Five simulations starting from $u$=0.2, 0.35, 0.5, 0.65, and 0.8 are plotted in different colors. Each simulation starts from an Erd\H{o}s-R\'{e}nyi graph with N=100,000 nodes and average degree $\lambda=4$. After initial transients, the fraction of discordant edges behaves as a function of the population of opinions. (Bottom) Similarly, the number of $0$-$1$-$0$ connected triples behaves as a function of $N_1/N$ after an initial transient (one simulation).}
  \label{fig:N1N010v3}
\end{figure}

\ms
To further explain the phrase ``quasi-stationary distributions'' in this context, we refer the reader to Figure \ref{fig:N1N010v3}. Let $N_1(t)$ be the number of vertices in state 1 at time $t$, $N_{01}(t)$ be the number of $0$-$1$ edges (that is, the number of edges connecting nodes $x$ and $y$ with $\xi(x)=0$, $\xi(y)=1$). Similarly, let $N_{abc}(t)$ be the number of connected triples $x$-$y$-$z$ with $\xi(x)=a$, $\xi(y)=b$, and $\xi(z)=c$. The top panel of Figure \ref{fig:N1N010v3} plots $N_{01}(t)/M$ versus $N_1(t)/N$ for five different simulations (with different initial densities, $u$) for $\alpha=0.5$. Note that in each case the simulation rapidly approaches a curve $\approx 1.710x(1-x) - 0.188$ and then diffuses along the curve until consensus is reached ($N_{01}=0$). At both of the possible consensus points on the curve, the fraction of the minority opinion is $\approx 0.12$, in accordance with the simulation in Figure \ref{fig:rewire_to_random_P1}.

The bottom panel of Figure \ref{fig:N1N010v3} similarly plots $N_{010}(t)/N$ versus $N_1(t)/N$ for $\alpha=0.5$ and $u=1/2$. Again the simulation rapidly approaches a curve (approximately cubic) and diffuses along it until consensus is reached. Since $N_{010}=0$ if $N_{01}=0$, and it is very unlikely that all $0$-$1$'s only occur in $0$-$1$-$1$ triples, the zeros of the cubic curve for $0$-$1$-$0$ and quadratic curve for $0$-$1$ coincide.

One can repeat the simulations in Figure \ref{fig:N1N010v3} for other network measurements, with the result that their values are similarly determined by the density $u(t)=N_1(t)/N$. This is somewhat analogous to a stationary distribution from equilibrium statistical mechanics---e.g., the Maxwell-Boltzmann distribution associating the velocity distribution with the temperature. We call our distributions quasi-stationary because our system is a finite state Markov chain, which will eventually reach one of its many absorbing states $N_{01}=0$, and hence there is no true stationary distribution. Nevertheless, an improved understanding of the system is obtained from these observations, displaying a fast dynamics rapidly converging to a family of neutrally-stable quasi-stationary distributions followed by slow, diffusive dynamics through the space local to the quasi-stationary distributions until consensus is reached.

To begin to explain the behavior of $\theta_t$ given in (\ref{WF}), note that when an edge is picked with two endpoints that differ, a rewiring will not change the number of 1's, while a voting event, which occurs with probability $(1-\alpha)$, will result in an increase or decrease of the number of 1's with equal probability. When $\theta_t=u$ the rate at which $0$-$1$ edges are chosen is equal to the expected fraction of $0$-$1$ edges under $\nu_{\alpha,u}$, which is $c_\alpha u(1-u) - b_\alpha$.

As shown in \cite{evo8}, the behaviors for the rewire-to-same model in terms of quasi-stationary distributions are very similar, but with small differences from the rewire-to-random model that yield fundamentally different consensus states. In rewire-to-same, there are quasi-stationary distributions $\nu'_{\alpha,u}$ under which the expected fraction of $0$-$1$ edges is $c'_\alpha u(1-u)$. Again the simulation comes rapidly to this curve and diffuses along it until consensus is reached. That is, unlike Figure \ref{fig:N1N010v3} (Top), the arches of quasi-stationary $N_{01}/M$ values versus $N_1/N$ maintain their zeros at $N_1/N=\{0,1\}$. Thus, for $\alpha < \alpha'_c(1/2)$, the minority fraction obtained at the consensus time is always $\approx 0$ for rewire-to-same.

\section{Multi-opinion models}
\label{sec:multi}

B\"ohme and Gross \cite{BG3} have studied the three-opinion version of the evolving voter model with rewire-to-same dynamics. In this case, the limiting behavior is complicated -- one may have partial fragmentation (1's split off rapidly from the 2's and 3's) in addition to full fragmentation and coexistence of the three opinions. See their Figures 3--5. As we will see in the present section, the behavior of the multi-opinion rewire-to-random model is much simpler because small groups of individuals with the same opinion will be drawn back into the giant component. We thus aim to extend the understanding of the two-opinion model behavior to larger numbers of opinions.

\begin{figure} 
  \centering
  \includegraphics[width=0.5\textwidth]{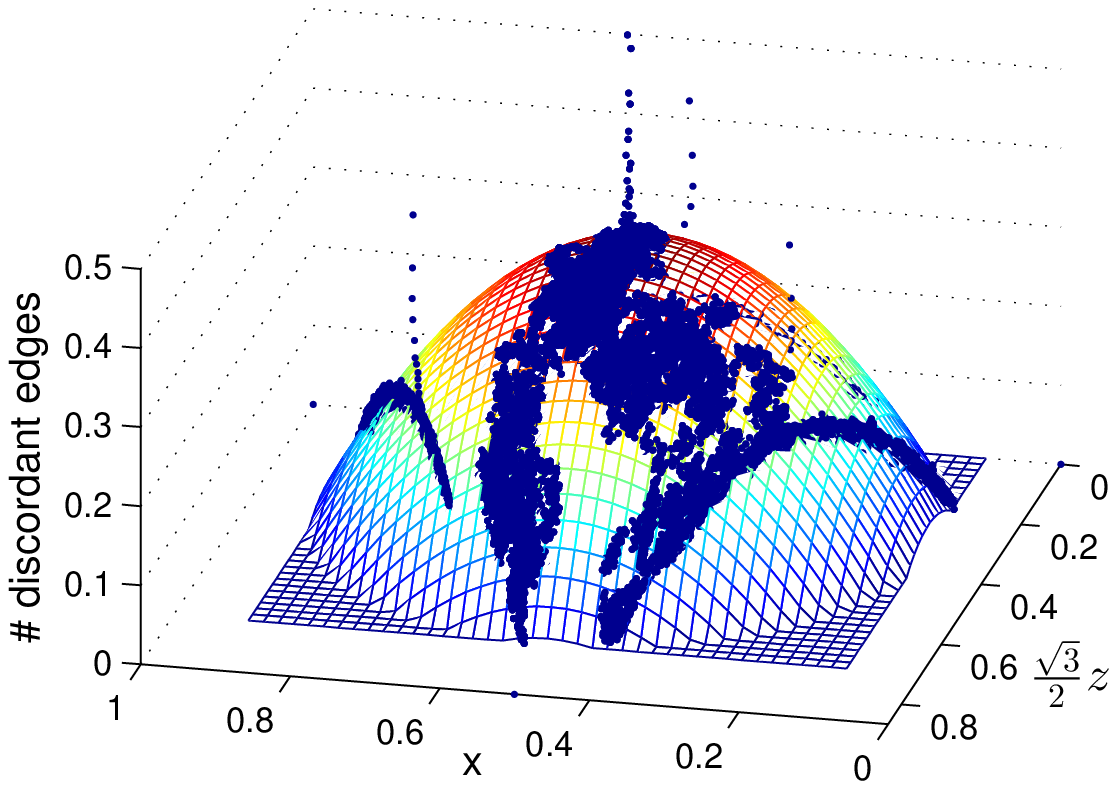}
  \includegraphics[width=0.5\textwidth]{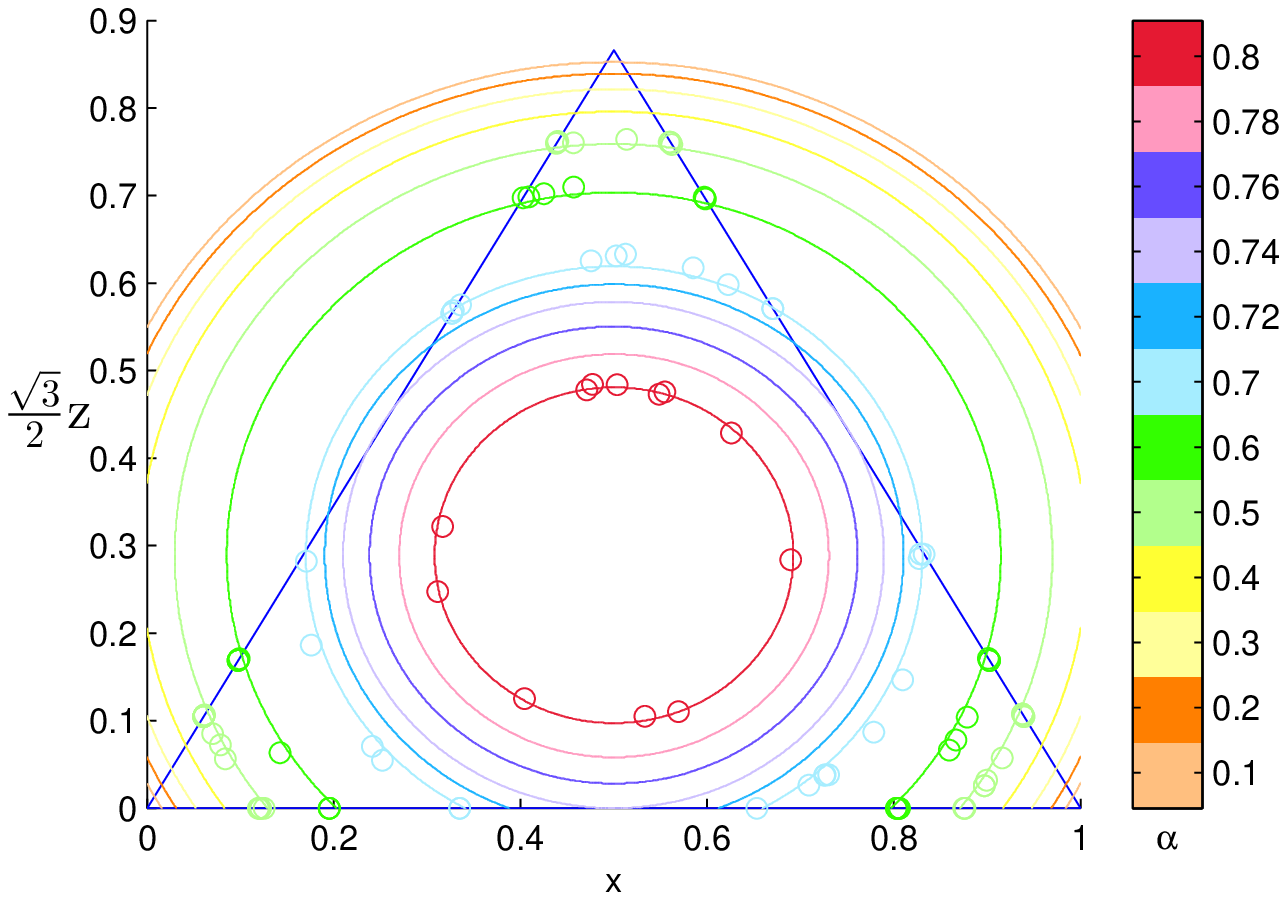}
  \caption{(Color Online) Top: plot of the fraction of discordant edges versus the population of opinions in barycentric coordinates for three opinions and $\alpha=0.5$. Multiple simulations corresponding to different initial densities are shown while each one starts from an Erd\H{o}s-R\'{e}nyi graph with N=10,000 nodes and average degree $\lambda=4$. Similar to the two-opinion case, the simulations quickly converge to a parabolic cap of quasi-stationary distributions. Bottom: top view of the parabolic caps of quasi-stationary distributions for $\alpha$=0.1,0.2,...,0.8. We fit the parabolic cap Eq. (\ref{quadcap}) to simulation data at various $\alpha$'s and then plot the level sets $E_uN_{\neq}=0$, which are the intersections of the parabolic caps with the $N_{\neq}=0$ plane, as the large circles with colors indicating values of $\alpha$.}
  \label{fig:levelset2}
\end{figure}

Consider now the $k$-opinion model in which voters are assigned independent initial opinions that are equal to $i$ with probability $u_i$. Let $u=(u_1,u_2,...,u_k)$ and let $N_{\neq}$ be the number of edges at which the endpoint opinions differ. When $k=3$, frequencies of the three types must lie in the triangle of possible values $\Delta = \{ u= (u_1,u_2,u_3) : u_i \ge 0, \sum_i u_i=1 \}$. To preserve symmetry, we draw $\Delta$ as an equilateral triangle in barycentric coordinates by mapping $(x,y,z) \to (x,z\sqrt{3}/2)$. The top panel in Figure \ref{fig:levelset2} plots $N_{\neq}(t)/M$ as a function of the opinion densities as the system evolves, generalizing the one-dimensional arch observed for $k=2$ to a two-dimensional cap for $k=3$.

Generalizing the parabolic form of the arch for $k=2$, we conjecture
\beq
E_{u} N_{\neq}/M = \frac{c_2(\alpha)}{2} \left( 1- \sum_{i=1}^k u_i^2 \right)  - c_0(\alpha).
\label{quadcap}
\eeq
As in the two opinion case, the simulated values come quickly to the surface and then diffuse along it. In some situations, one opinion is lost before consensus occurs and the evolution reduces to that for the two opinion case. However, in one of the simulations shown, the realization ending with $x\approx 0.5$, all three opinions persist until the end.

The picture is somewhat easier to understand if we look at the cap from a top view, where the $E_{u} N_{\neq}=0$ level sets for different $\alpha$ are observed to be circles. In the bottom panel of Figure \ref{fig:levelset2} we plot the $E_{u} N_{\neq}=0$ circles for different $\alpha$'s fitted from simulation data using Eq. (\ref{quadcap}) as well as the consensus opinion frequencies from the simulations (indicated by small circle data points). The two agree with each other up to small stochastic fluctuations. The size of the $E_{u} N_{\neq}=0$ level set then dictates different consensus state properties. For example, the circle corresponding to $\alpha=0.5$  intersects $\Delta$ in three disconnected arcs. As $\alpha$ increases, the radius of the $E_{u} N_{\neq}=0$ level set decreases. When $\alpha>\alpha_c(1/2)$, the critical value of the two opinion model, the circle $E_{u} N_{\neq}=0$ falls fully inside the triangle, so an initial condition including all three opinions will continue to demonstrate all three opinions at consensus. For example, the small circles around the innermost circle give the ending frequencies for several simulations for $\alpha=0.8$. If the initial frequencies fall within the $E_{u} N_{\neq}=0$ circle, then the model will quickly relax to the quasi-stationary distributions above the circle and then diffuse along the cap until consensus is reached at some $E_{u} N_{\neq}=0$ point. If instead the initial frequencies $u$ fall outside the $E_{u} N_{\neq}=0$ circle---that is, for $\alpha$ above the phase transition point $\alpha_3(u)$---the consensus time jumps from $O(N^2)$ to $O(N\log N)$, similar to $\alpha_c(u)$ for the two-opinion model, with the final opinion frequencies essentially the same as the initial $u$. What is new in this case is that when starting with three opinions and $\alpha_c(u)<\alpha<\alpha_3(u)\leq\alpha_3(\{\frac{1}{3},\frac{1}{3},\frac{1}{3}\})$, the system always ends up with three distinct opinions.

For $k>3$, our simulation results indicate the same type of behavior as the system evolves. We define $\alpha_k$ to be the largest $\alpha$ for which consensus takes $O(N^2)$ updates when we start with $k$ opinions with density $1/k$ for each opinion. Then as $k\rightarrow\infty$ the multi-opinion model has infinitely many phase transitions. When $\alpha_k<\alpha<\alpha_{k+1}$, consensus occurs after $O(N\log N)$ steps if we start with $k$ opinions, while if we start with $k+1$ equally likely opinions the system quickly converges to a quasi-stationary distribution and diffuses until consensus occurs after $O(N^2)$ updates and there will always be $k+1$ opinions present at the end. The associated picture is the natural dimensional extension of the relationship between the $k=2$ and $k=3$ models: just as $\alpha_2=\alpha_c(1/2)$ corresponds to the point at which the $E_{u} N_{\neq}=0$ circle for $k=3$ is the inscribed circle within the $\Delta$ triangle, $\alpha_3$ corresponds to the point at which the $E_{u} N_{\neq}=0$ circle reaches zero radius---that is, the point at which the $E_{u} N_{\neq}=0$ sphere for $k=4$ has become the inscribed sphere within the corresponding barycentric tetrahedron.

\section{Quantitative characterization of quasi-stationary distributions}
\label{sec:quant}

For each $k$ we simulate our multi-opinion rewire-to-random model starting from $k$ opinions with each opinion taking $1/k$ fraction of nodes at random for a wide range of $\alpha$'s. Generalizing the picture of the one-dimensional arch for $k=2$ and the two-dimensional cap for $k=3$, the number of discordant edges as a function of frequencies conjectured in Eq.~(\ref{quadcap}) is a co-dimension 1 hypersurface characterizing the quasi-stationary states, and the behavior of the equal-initial-populations case will allow us to describe this surface, thereby characterizing behaviors for general initial populations.

\begin{figure*}
  \begin{minipage}{0.5\textwidth}
    \includegraphics[width=\textwidth]{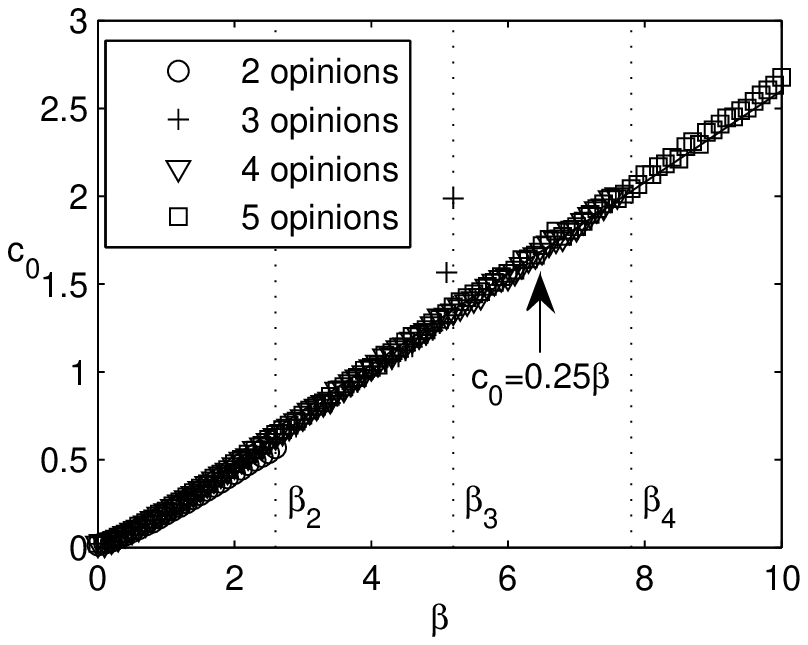}
  \end{minipage}%
  \begin{minipage}{0.5\textwidth}
    \includegraphics[width=\textwidth]{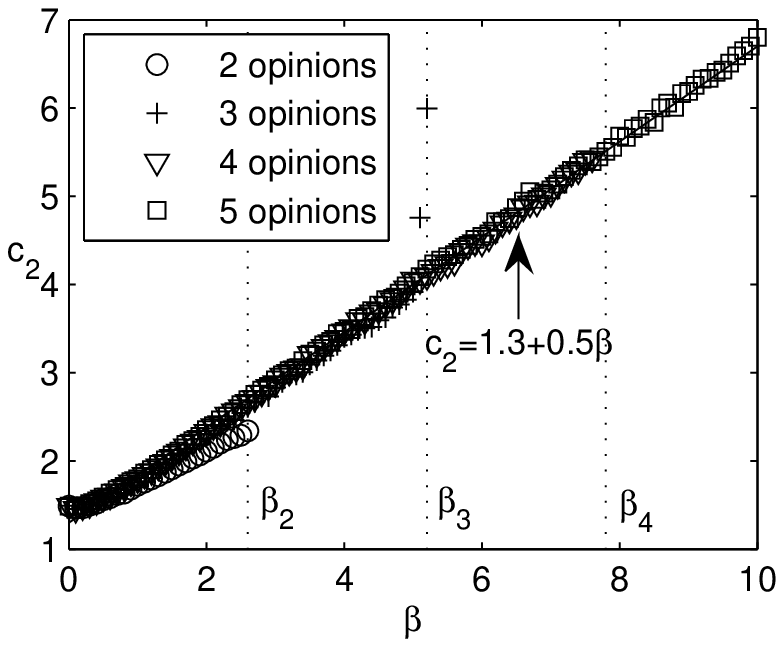}
  \end{minipage}
  \caption{Coefficient $c_0(\beta)$ (left) and $c_2(\beta)$ (right) in Eq.~(\ref{quadcap}) for models with multiple opinions. Each value of the coefficients is obtained by fitting Eq.~(\ref{quadcap}) to multiple simulations starting from Erd\H{o}s-R\'{e}nyi graphs with N=100,000 nodes and average degree $4$. The fitting error is very small ($R^2\approx0.99$) except for $\beta$ close to the critical values.}\label{fig:c_2}
\end{figure*}

First the critical $\alpha_k$'s are identified when the slow diffusion of $N_{\neq}$ cannot be observed for the first time as $\alpha$ increases from $0$ to $1$. Then we fit $N_{\neq}(t)/M$ to $u_i(t)=N_i(t)/N$ ($i=1,...,k$) using Eq.~(\ref{quadcap}) at every $\alpha$ up to $\alpha_k$, and plot the fitted coefficients $c_0$ and $c_2$ against $\beta=\alpha/(1-\alpha)$ in Figure \ref{fig:c_2}. Remarkably, the coefficients in (\ref{quadcap}) appear to be well approximated by linear functions of $\beta = \alpha/(1-\alpha)$. The graphs shows some curvature near $\beta=0$, which may be caused by the fact that $\beta=0$ ($\alpha=0$) corresponds to a voter model without evolution of the underlying network. In the rest of the paper, we will work with $\beta$ for simplicity. Naturally, critical points $\alpha_k$ translate to $\beta_k=\alpha_k/(1-\alpha_k)$.

\begin{figure*}
  \begin{minipage}{0.5\textwidth}
    \includegraphics[width=\textwidth]{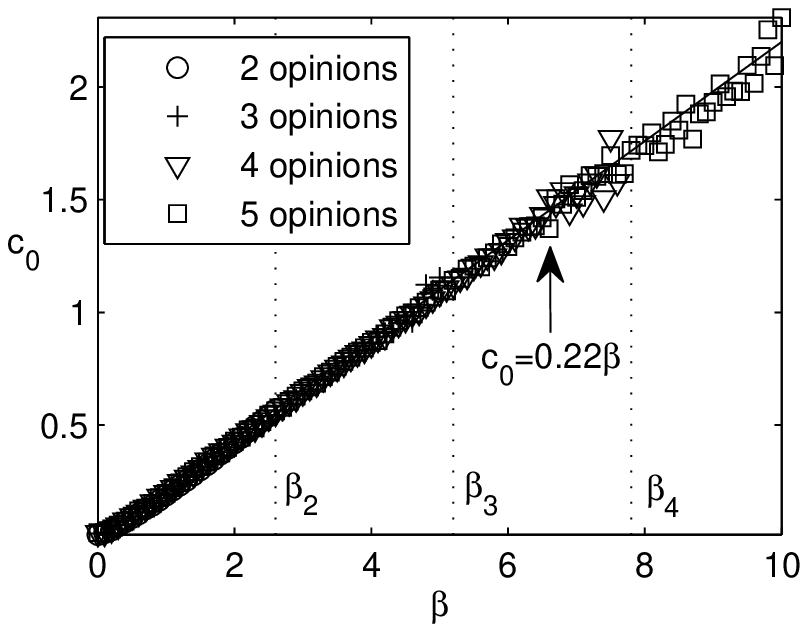}
  \end{minipage}%
  \begin{minipage}{0.5\textwidth}
    \includegraphics[width=\textwidth]{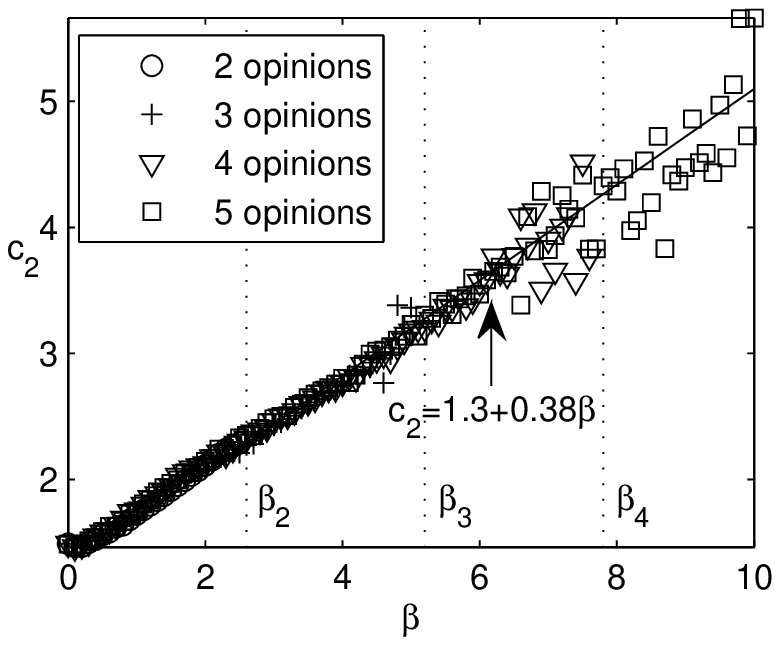}\\
  \end{minipage}
  \caption{Coefficients $c_0(\beta)$ (left) and $c_2(\beta)$ (right) in Eq.~(\ref{kthcap}) for models with multiple opinions. Each value of the coefficients is obtained by fitting Eq.~(\ref{kthcap}) to the same data as in Figure \ref{fig:c_2}.}\label{fig:c_2_high}
\end{figure*}

The fitted coefficients from the 2-opinion model deviate slightly from those fitted from higher-order models, which implies that Eq.~(\ref{quadcap}) is not universal for the multi-opinion model and higher-order terms are possible. However, while the discrepancy between the fitted coefficients of the 2-opinion model and those of the 3-opinion one is apparent, difference between fitted coefficients of higher-order models is negligible, which implies that the inclusion of higher-order terms beyond the 3rd would not make significant changes to the equation. To probe the effect of higher-order terms we introduce terms up to $k$th order for $k$ opinions. Noting $(\sum_i u_i)^2=1$, Eq.~(\ref{quadcap}) is equivalent to:
\beq
E_u N_{\neq}/M = - c_0(\alpha) + c_2(\alpha)\sum_{i,j=1;i>j}^k u_i u_j .
\eeq
Given the symmetry of the system in $u_i$'s, the only possible choice in degree-k polynomials is:
\begin{eqnarray}
E_u N_{\neq}/M &=& -c_0(\alpha) + c_2(\alpha)\sum_{\{i_1,i_2\}\in\mathcal{A}_2} u_{i_1}u_{i_2} \nonumber\\ &&+c_3(\alpha)\sum_{\{i_1,i_2,i_3\}\in\mathcal{A}_3} u_{i_1}u_{i_2}u_{i_3} + \cdots \nonumber\\ &&+c_k(\alpha)\sum_{\{i_1,\cdots,i_k\}\in\mathcal{A}_k} u_{i_1}u_{i_2}\cdots u_{i_k},
\label{kthcap}
\end{eqnarray}
where $\mathcal{A}_i$ is the collection of all $i$-element subsets of $\{1,2,...,k\}$. Using the same simulation data as above, we refit $N_{\neq}(t)/M$ to $u_i(t)$'s ($i=1,...,k$) according to the generalized formula Eq. (\ref{kthcap}) and plot the fitted coefficients $c_0$ and $c_2$ against $\beta$ in Figure \ref{fig:c_2_high}. Fitting diagnostics suggest that higher-order terms are significant from zero (with $p\text{-value} < 10^{-4}$) and it can be seen that those terms explain the inconsistency between fitted coefficients of different models in Figures \ref{fig:c_2}. However, the difference between the two fitted functions of Eq. (\ref{quadcap}) and Eq. (\ref{kthcap}) is actually small ($\approx.1$ in $L^2$-norm) and thus higher-order terms are small corrections to the hyper-surface Eq. (\ref{quadcap}).

Values of the coefficients $c_i(\beta)$ for the three opinion model near its critical value $\beta_3\approx5.2$ show some scatter, but this is to be expected since the surface is very small at this point. Values for the four opinion model appear to become more difficult to fit prior to $\beta_4$ since $E_u N_{\neq}=0$ is a three-dimensional hyper-surface in four-dimensional space, so much more data is required to get reliable estimates of coefficients.

\begin{figure}
  \includegraphics[width=0.5\textwidth]{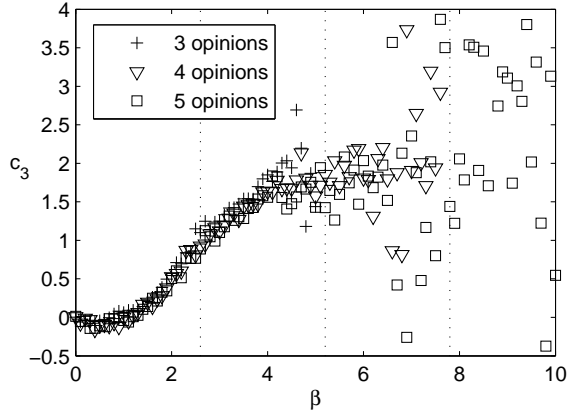}\\
  \caption{Coefficient $c_3(\beta)$ in Eq.~(\ref{kthcap}) for models with multiple opinions. Each value of $c_3(\beta)$ is obtained by fitting Eq.~(\ref{kthcap}) to the same data in Figure \ref{fig:c_2}.}\label{fig:c_3}
\end{figure}

As is visually apparent in Figure \ref{fig:c_2_high}, the coefficients $c_0$ and $c_2$ for the first two terms in Eq.~(\ref{kthcap}) are well approximated by linear functions, with best fits $c_0(\beta)\approx 0.22\beta$ and $c_2(\beta)\approx 1.3 + 0.38\beta$, while coefficients for higher-order terms are not linear in $\beta$ (e.g., see Figure \ref{fig:c_3} for $c_3(\beta)$). For comparison, the best fits for $c_0$ and $c_2$ in Eq.~(\ref{quadcap}) (as in Figure \ref{fig:c_2}) are
\beq
c_2(\beta) \approx 1.3 + 0.5\beta, \qquad c_0(\beta) \approx 0.25\beta.
\label{coeff}
\eeq
Since Eq. (\ref{quadcap}) well approximate the higher-order hyper-surface Eq. (\ref{kthcap}), its simple form can be used to estimate the critical points for phase transitions. Combining (\ref{quadcap}) and (\ref{coeff}), and then solving
\[
( 0.65 + 0.25\beta) ( 1- k(1/k)^2 )  - 0.25\beta = 0
\]
gives
\[
\beta_k=2.6(k-1)\,.
\]
which agrees with the critical $\beta_k$'s identified when the slow diffusion of $N_{\neq}$ cannot be observed in simulations as $\beta$ increases.

\section{Conclusion}
\label{sec:conclusion}

Our multi-opinion voter model has infinitely many phase transitions. When $\beta_k < \beta < \beta_{k+1}$, consensus occurs rapidly when we start with $k$ opinions, while if we start with $k+1$ equally likely opinions there will always be $k+1$ opinions present at the end. To a good approximation $\beta_k = 2.6(k-1)$, but the departures from linearity in the plots of $c_2(\beta)$ and $c_0(\beta)$ suggest that this result is not exact. However, formulas for various quantities associated with this model are close to polynomials, so an exact solution may be possible.

More complicated rewiring rules might also be considered, particularly if they maintained high clustering or other global macroscopic properties. An even more complete understanding of the present rewiring system would help motivate similar investigations for other rewiring rules.

\end{document}